\documentclass[manuscript]{aastex}
\usepackage{epsfig}
\usepackage{graphicx}
\usepackage{txfonts}
\usepackage{natbib}
\def\arcmin{\hbox{$^\prime$}}
\def\arcsec{\hbox{$^{\prime\prime}$}}
\def\farcs{\hbox{$.\!\!^{\prime\prime}$}}

\slugcomment{To appear in Astrophysical Journal}

\shorttitle{A ring/disk/outflow system associated with W51 North: 
a very massive star in the making}
\shortauthors{Zapata et al.}

\begin{document}


\title{A ring/disk/outflow system associated with W51 North: 
\\ a very massive star in the making}


\author{Luis A. Zapata\altaffilmark{1}, Paul T. P. Ho\altaffilmark{2,3}, 
 Peter Schilke\altaffilmark{1}, Luis F. Rodr\'\i guez\altaffilmark{4},
 Karl Menten\altaffilmark{1}, \\ Aina Palau\altaffilmark{5}, and Robin
 Garrod\altaffilmark{6}}


\altaffiltext{1}{Max-Planck-Institut f\"{u}r Radioastronomie, Auf dem H\"{u}gel
69,53121, Bonn, Germany}
\altaffiltext{2}{Harvard-Smithsonian Center for Astrophysics, 60 Garden Street,
Cambridge, MA 02138, USA}
\altaffiltext{3}{Academia Sinica Institute of Astronomy and Astrophysics,
Taipei, Taiwan}
\altaffiltext{4}{Centro de Radioastronom\'\i a y Astrof\'\i sica,
 UNAM, Apdo. Postal 3-72 (Xangari), 58089 Morelia, Michoac\'an,
 M\'exico} \altaffiltext{5}{LAEX, Centro de Astrobiolog{\'i}a (CAB,
 INTA-CSIC), LAEFF, P.O. Box 78, E-28691 Villanueva de la Ca\~nada,
 Madrid, Spain} \altaffiltext{6}{Department of Astronomy, Cornell
 University, 106 Space Sciences Building Ithaca NY 14853 USA}


\begin{abstract}
Sensitive and high angular resolution ($\sim$ 0.4\arcsec)
SO$_2$[22$_{2,20}$ $\rightarrow$ 22$_{1,21}$] and SiO[5$\rightarrow$4]
line and 1.3 and 7 mm continuum observations made with the
Submillimeter Array (SMA) and the Very Large Array (VLA) towards the
young massive cluster W51 IRS2 are presented.  We report the presence
of a large (of about 3000 AU) and massive (40 M$_\odot$) dusty
circumstellar disk and a hot gas molecular ring around a
high-mass protostar 
or a compact small stellar system associated with W51 North. 
The simultaneous observations of the silicon monoxide molecule, 
an outflow gas tracer, further revealed a massive (200 M$_\odot$) 
and collimated ($\sim$
14$^\circ$) outflow nearly perpendicular to the dusty and molecular
structures suggesting thus the presence of a single very massive protostar
with a bolometric luminosity of more than 10$^5$ L$_\odot$.
A molecular hybrid LTE model of a
Keplerian and infalling ring with an inner cavity  and 
a central stellar mass of more than 60 M$_\odot$ agrees well with the
SO$_2$[22$_{2,20}$ $\rightarrow$ 22$_{1,21}$] line observations.
Finally, these results suggest that mechanisms, such as mergers 
of low- and intermediate- mass stars,
might be not necessary for forming very massive stars.  
\end{abstract}


\keywords{ stars: pre-main sequence -- ISM: individual: (W51 IRS2) 
-- ISM: Molecules -- radio continuum }



\section{Introduction}
In recent years a small group of candidate accreting disks in
high-mass protostars has been reported in the literature \citep[and
references therein]{cesaronietal2006}, but with luminosities typical
of B-type main-sequence stars, that is, stars with masses $\leq$ 25
M$_\odot$. Around more luminous and massive objects (presumed O-type
stars) there have been no clear evidence for accretion disks, only
gravitational unstable and large rotating molecular structures have
been found. These molecular structures (``{\it toroids}'') are
infalling and thus accreting fresh gas material to a central cluster
of young massive protostars \citep{sollinsetal2005,
beltranetal2005,beltranetal2006} or maybe a single 40 M$_\odot$
protostar \citep{sandelletal2003,be2008}. The sizes and masses of the
rotating toroids are about 2-3 $\times$ 10$^4$ AU and 100-400
M$_\odot$, respectively.

Located at 5-8 kpc away in the Sagittarius spiral arm
\citep{genzeletal1981,imaietal2002,Schnepsetal1981} and with a total
bolometric luminosity of about 3 $\times $ 10$^6$ $L_{\odot}$, the
W51-IRS2 region is one of the most luminous massive star forming
regions in our Galaxy \citep{Eri1980}. 

Recently however it has been estimated a distance of 2.3 $\pm$ 0.3 kpc to
W51-IRS2 using spectroscopic parallaxes of OB stars \citep{Fig2008}.
But, in other hand, a lower limit of 5 kpc by triangulation using Very Long
Baseline Array observations of methanol masers located in W51 IRS2 was
found by \citet{xu2008}, a very similar value to those found earlier
by the statistical H$_2$O maser parallax 
\citep{genzeletal1981,Schnepsetal1981, imaietal2002}. 
In this study, we adopt a distance of 6 kpc for the cluster.

This region might contain about 30 O-type zero-age main-sequence stars
with excess emission at the infrared and (sub)millimeter wavelengths,
see for an example \citet{bar2008}. In the W51 IRS2 cluster there are
two highly obscured massive sources called {\it W51 North} and {\it
W51 d2} that seem to be the youngest objects in the cluster and that
exhibit strong emission from many complex molecules
\citep{ho1983,rud1990,zha1997,zha1998,soll2004,zap2008}. In
particular, toward the W51 North object very strong maser emission
from various species {\it e.g.}, hydroxyl (OH), silicon monoxide
(SiO), and water (H$_2$O) has been detected
\citep{Schnepsetal1981, gau1987, mor1992}.  This type of emission is
associated with the formation of the high-mass stars. Observations of
the proper motions of the H$_2$O masers, in addition, revealed the
presence of a remarkable compact ($\sim$ 1-2\arcsec )
southeast-northwest outflow emanating from this protostar
\citep{Schnepsetal1981,ein2002,imaietal2002}.
Furthermore, very high spatial resolution VLA and VLBA radio 
observations revealed that the SiO maser emission traces the innermost parts 
of the outflow ejected from W51 North \citep{ein2002}. Finally, observations of
the cyanogen (CN) molecule showed that the molecular gas is falling
into this central protostar with a mass accretion 
rate $\sim$ 10$^{-3}$ $M_\odot$ yr$^{-1}$ \citep{zap2008}.

\begin{figure*}[!ht]
\begin{center}\hspace{0cm} 
\includegraphics[scale=0.18, angle=0]{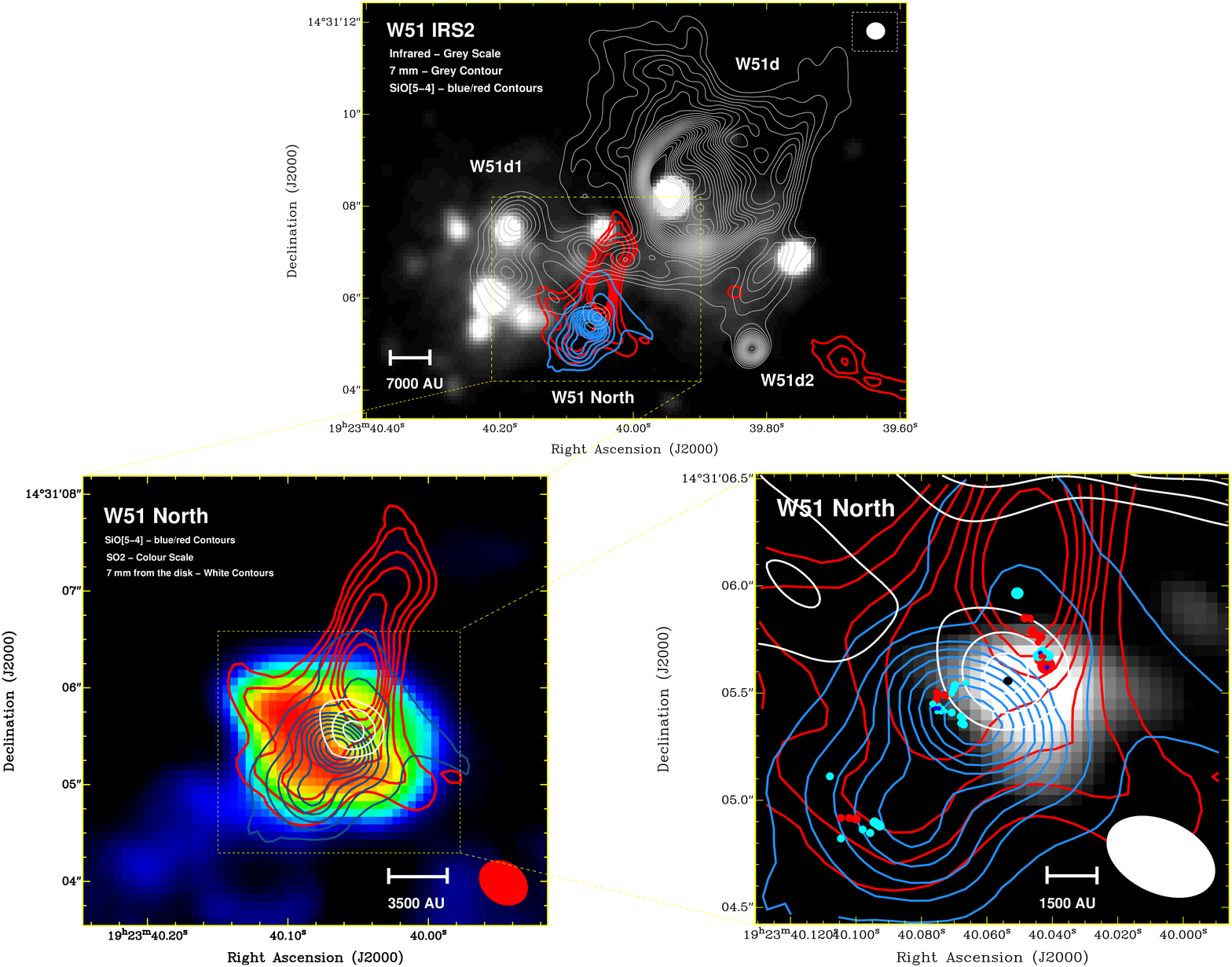}
\caption{\scriptsize {\bf \it Upper Panel:} Overlay of an archival VLT H-band
image (grey scale), a VLA 7 mm continuum image (grey contours) and a
SMA SiO[5$\rightarrow$4] moment zero line image (blue and red
contours) of the young massive cluster W51 IRS2. Blue and red contours
correspond respectively to the blue- and red-shifted gas that emanates
from the obscured protostar W51 North. The 7 mm contours are from 10\%
to 90\% with steps of 3\% of the peak of the continuum emission, the
peak is 100 mJy beam$^{-1}$. The synthesized beam of the 7 mm contour
image is 0.46$''$ $\times$ 0.43$''$ with a P.A. of -53.6$^\circ$ and
it is shown in the upper right corner of the box. The
SiO[5$\rightarrow$4] line contours are from 20\% to 90\% with steps of
10\% of the peak of the line emission, the peak is 5 Jy beam$^{-1}$ km
s$^{-1}$ for the red-shifted emission and 8 Jy beam$^{-1}$ km s$^{-1}$
for the blue-shifted. The integrated velocity range for the
blueshifted gas is from +20 to +58 km s$^{-1}$ and that for the
redshifted gas is from +60 to +95 km s$^{-1}$.  {\bf \it Left Lower
Panel:} A zoom into the W51 North zone. Overlay of the
SO$_2$[22$_{2,20}$ $\rightarrow$ 22$_{1,21}$] moment zero line
emission (color image), the 7 mm continuum emission from the disk
(white contours) and the SiO[5$\rightarrow$4] moment zero line
emission.  The integrated velocity range for the SO$_2$[22$_{2,20}$
$\rightarrow$ 22$_{1,21}$] is from +50 to +70 km s$^{-1}$.  The
synthesized beam of the SiO[5$\rightarrow$4] contour image is 0.58$''$
$\times$ 0.43$''$ with a P.A. of 54.8$^\circ$ and it is shown in the
bottom right corner of the box. {\bf \it Right Lower Panel:} A zoom
into the dusty disk and the outflow.  The grey scale image shows the
1.3 mm line-free continuum emission overlaid with the 7 mm continuum
emission (white contours) and SiO[5$\rightarrow$4] line emission
(blue/red contours).  The blue and red dots mark the position of the
blue- and red-shifted water maser spots, respectively, reported by
\citet{ein2002}. The synthesized beam of the 1.3 mm continuum image is
0.50$''$ $\times$ 0.35$''$ with a P.A. of 54.8$^\circ$ and it is shown
in the bottom right corner of the box. For the 1.3 mm image, we used
the ROBUST parameter of the INVERT task of MIRIAD set to -2, which
corresponds to uniform weighting to achieve the maximum angular
resolution. The black dot marks the position of SiO maser source given
by \citet{ein2002}, ($\alpha$, $\delta$)$_{J2000}$= (19$^h$ 23$^m$
40\farcs055, 14$^\circ$ 31\arcmin ~5\farcs59).}
\end{center}
\end{figure*}

\section{Observations}

In order to study the very young massive object {\it W51 North}, we
carried out observations of the SO$_2$[22$_{2,20}$ $\rightarrow$
22$_{1,21}$] (sulfur dioxide) and SiO[5$\rightarrow$4] (silicon
monoxide) molecular lines and 1.3 mm continuum with the Submillimeter
Array. We also obtained 7 mm continuum data with the Very Large
Array. Both observations had high angular resolution ($\sim$
0.4$''$). 

The SO$_2$ molecule is a good high-density gas tracer 
\citep[see][]{leu07,jimena2007}, while
the SiO, is an excellent tracer for molecular outflows.\footnote{ 
The production of the silicon monoxide (SiO) molecule is mainly attributed 
to the destruction of the dust particles in strong shocks \citep{sch97,van98}.} 
The SMA SiO[5$\rightarrow$ 4], SO$_2$[22$_{2,20}$ $\rightarrow$ 22$_{1,21}$]
and 1.3 mm observations were made simultaneously on 2008 January 17,
and the VLA 7 mm observations were made on 2008 April 22.  At that
time, the SMA was in its very extended configuration, while the VLA
was in its C configuration. The phase reference center of both VLA and
SMA observations was R.A.=19$^h$ 23$^m$ 40\farcs055, decl.= 14$^\circ$
31\arcmin ~5\farcs59 (J2000.0). 

For the 7 mm observations, we integrated on-source for a total of approximately 5 hr
using the fast-switching mode with a cycle of 120 s.
The observations were made using the continuum mode,
with a total bandwidth of 100 MHz. 
The central frequency observed was 43.34 GHz.
The absolute amplitude calibrator was 1331+305 
(with an adopted flux density of 1.45 Jy) 
and the phase calibrator was 1924+156 (with a bootstrapped flux density of 0.65 
$\pm$ 0.01 Jy). 

For the 1 mm observations, the zenith opacity ($\tau_{230 GHz}$), 
measured with the NRAO tipping
radiometer located at the Caltech Submillimeter Observatory, was
$\sim$ 0.15, indicating excellent weather conditions (for this
frequency) during the observations.  Observations of Uranus provided
the absolute scale for the flux density calibration.  Phase and
amplitude calibrators were the quasars 1925+211 and 2035+109, with
measured flux densities of 0.7 $\pm$ 0.1 and 0.5 $\pm$ 0.1 Jy,
respectively.  

The SO$_2$[22$_{2,20}$ $\rightarrow$ 22$_{1,21}$] and
SiO[5$\rightarrow$4] transitions were detected in the lower side band
(LSB) of the SMA at a frequency of 216.643 and 217.104 GHz,
respectively.  The full bandwidth of the SMA correlator is 4 GHz (2
GHz in each band). The SMA digital correlator was configured in 24
spectral windows (``chunks'') of 104 MHz each, with 256 channels
distributed over each spectral window, providing a resolution of 0.40
MHz (0.58 km s$^{-1}$) per channel.
Further technical descriptions of the SMA and its
calibration schemes can be found in \citet{ho2004}.

The data were edited, calibrated and
imaged using the programs MIR, AIPS, MIRIAD and KARMA.
In both bands, we used the ROBUST parameter set to 0 to obtain an optimal compromise
between sensitivity and angular resolution. The 7mm data were self-calibrated in phase.
For the 7 mm observations, the continuum image rms noise was 0.9 mJy beam$^{-1}$ at an angular resolution
of $0\rlap.{''}46$ $\times$ $0\rlap.{''}43$ with a P.A. = -53.6$^\circ$.
For the 1 mm observations, the continuum image rms noise was 6 mJy beam$^{-1}$ 
at an angular resolution of $0\rlap.{''}58$ $\times$ $0\rlap.{''}43$ with a P.A. = 54.8$^\circ$.
The line image rms noise was about 80 mJy beam$^{-1}$.




\begin{deluxetable}{lccccccc}
\tabletypesize{\scriptsize }
\tablecaption{Physical Parameters of the Molecular Ring and Circumstellar Disk}
\tablewidth{0pt}
\tablehead{
\colhead{} &
 \multicolumn{2}{c}{Position} & \colhead{} & \colhead{} & \colhead{} &
 \colhead{} &\\ \cline{2-3} Wave./Mol.  & $\alpha_{2000}$ &
 $\delta_{2000}$ & Flux Density& Deconvolved Size & P.A.  & Gas Mass &
 Dyn. Mass\\ & [$h$ $m$ $s$] & [$\circ$ $\prime$ $\prime\prime$] &
 [mJy] & [arcsec] & [Degree] & [M$_\odot$] & [M$_\odot$] }
\startdata
7 mm & 19 23 40.057 & 14 31 05.67 & 17 $\pm$ 3 & 0.58$\pm$0.02
$\times$ 0.27$\pm$0.02 & 70$\pm$6 & -- & --\\ 1.3 mm & 19 23 40.045 &
14 31 05.48 & 1620 $\pm$ 160 & 0.72$\pm$0.03 $\times$ 0.57$\pm$0.03 &
90$\pm$40 & 40 & --\\ SO$_2$ & 19 23 40.074 & 14 31 05.44 &
7$\times$10$^4$ $\pm$ 500$^1$ & 1.8$\pm$0.1 $\times$ 1.2$\pm$0.1 &
72$\pm$2 & -- & $\geq$ 100$^2$
\enddata
\tablecomments{ 1: The units are mJy Beam$^{-1}$ km
 s$^{-1}$.  2: The dynamical mass was obtained from our model.}.
\end{deluxetable}

\begin{figure}[!ht]
\begin{center} 
\includegraphics[scale=0.6, angle=0]{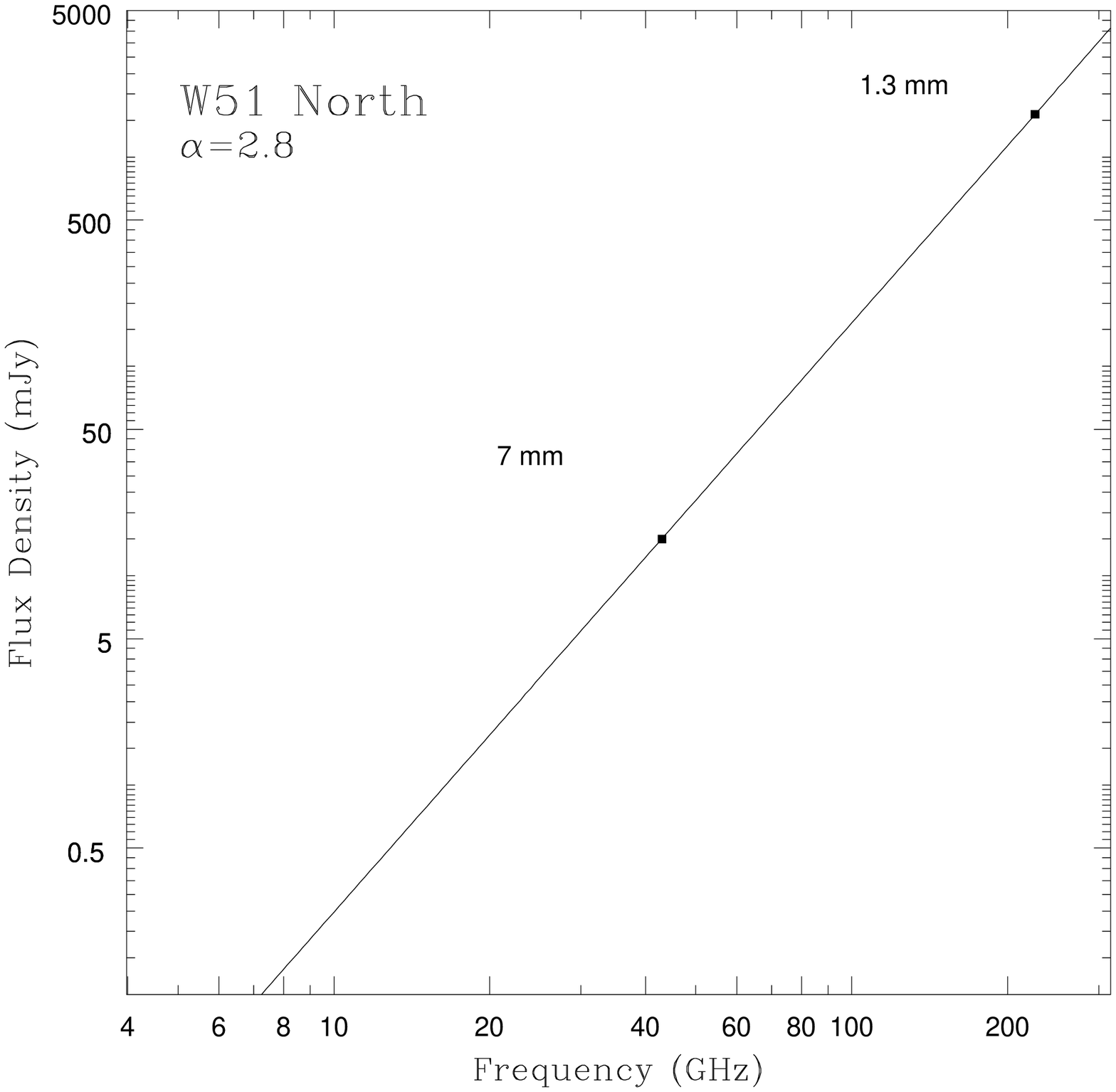}
\caption{\scriptsize SED for the source W51 North combining 
7 and 1.3 mm continuum data.  
The respective error bars were smaller than the squares and
are not presented. The line is a least-squares power-law fit 
(of the form S$_\nu \propto
\nu^\alpha$) to the spectrum.}
\end{center}
\end{figure}

\section{Results and Discussion}

\subsection{A massive and large dusty disk}

At a wavelength of 7 mm (see Figure 1), we detected strong free-free
continuum emission arising from the HII regions associated with the
neighboring young massive ZAMS stars located in the W51 IRS2
cluster. This emission has already been mapped at centimeter
wavelengths by many authors \citep{gau1987, ein2002, Meh1994, la2007}.
The radio images presented by \citet{la2007} were obtained from
\citet{Meh1994}.  However, in our map, we detected for the first time
a faint and compact millimeter source associated with the obscured
high-mass protostar located in W51 North. This 7 mm continuum source
is the counterpart of the dusty sources associated with large
envelopes, and reported at 2 and 1.3 mm by \citet{zha1998,zap2008},
respectively.

\begin{figure*}[!ht]
\begin{center} 
\includegraphics[scale=0.8, angle=0]{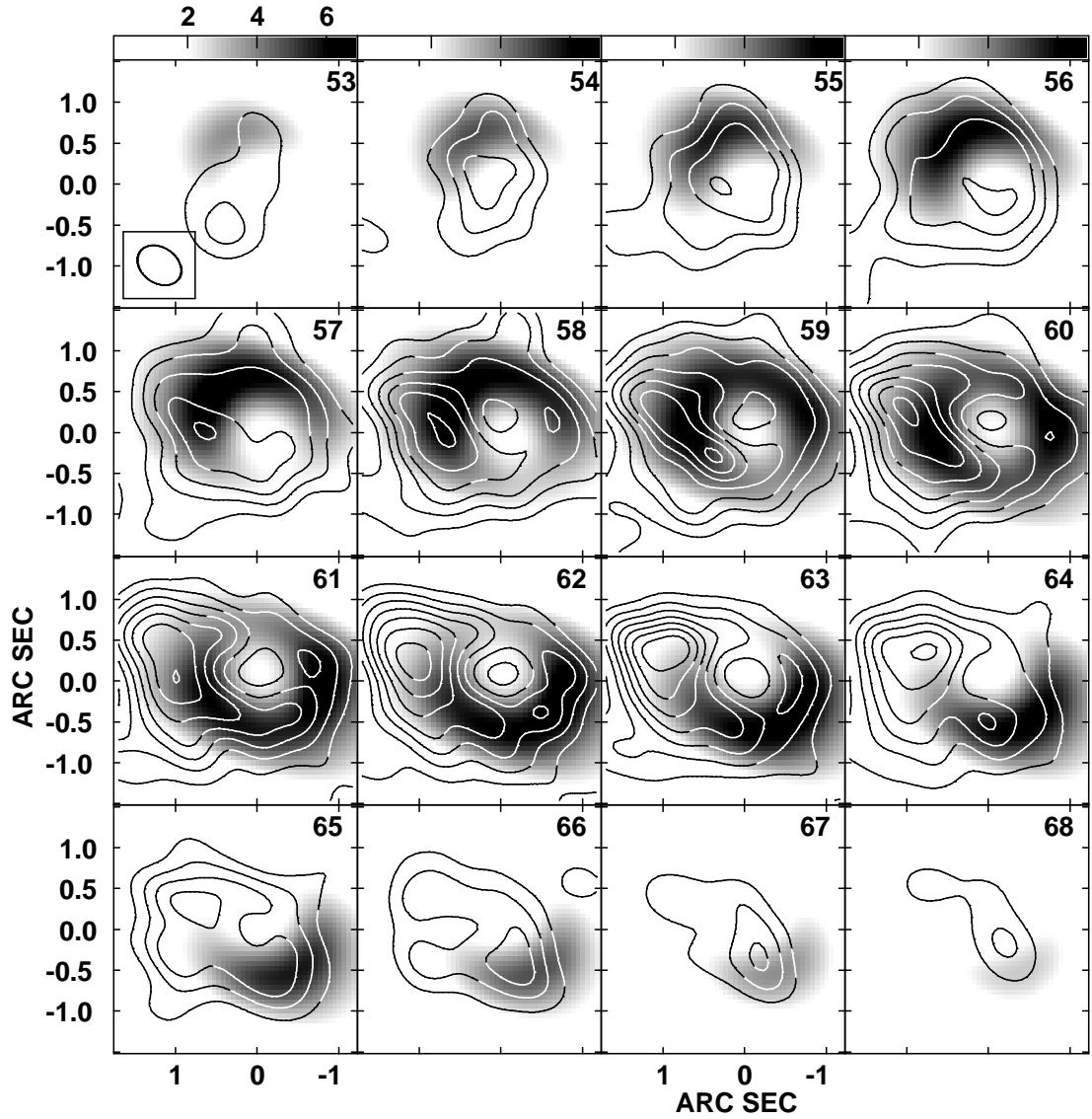}
\caption{
\scriptsize Velocity-channel image of the SO$_2$[22$_{2,20}$ $\rightarrow$  
22$_{1,21}$]
thermal emission from the ring (contours) overlaid with our synthetic LTE
model (greysacle). The spectral resolution was smoothed to velocity
bins of 1 km s$^{-1}$. The central velocity is indicated (in km
s$^{-1}$) in the top rigth-hand corner of each panel. The systemic
velocity of the ambient molecular cloud is about 60.0 km s$^{-1}$. The
contours are 10\% to 90\% with a step of 10\% of the peak flux.
The scale bar indicates the peak flux of the molecular emission in T$_b$.}
\end{center}
\end{figure*}

In Figure 2, we have constructed the spectral energy distribution
(SED) for this millimeter source from the centimeter to millimeter
wavelengths, combining only data presented here.  
The angular resolution of two different
observations are quite similar, of about 0.4$''$. From these data, we
estimated a spectral index of $\alpha$ = 2.8. This suggest that
the emission at these wavelengths likely arises from a dusty disk. 
A larger envelope has been already mapped at scales 10$^4$ AU as
mentioned above. 
The  hypothesis of a flattened circumstellar disk 
is supported by the fact that the mm source is 
resolved at wavelength 7 mm (see Table 1) and
show a modest east-west elongation, nearly perpendicular to the
orientation of the molecular outflow that will discussed on the
following sections.  At 1.3 mm the orientation is not so well
determined. 

We noted from Figure 16 of \citet{Meh1994} there is no strong emission
from the recombination lines H92$\alpha$ and He92$\alpha$ toward the
mm source, both lines are associated with the compact HII regions W51d
and/or d1.  The non detection of recombination lines emission from the
mm source is consistent with our interpretation. 
Taking the values of the deconvolved
sizes and the flux densities for the continuum source at 7 mm and 1.3
mm, we estimated a brightness temperature 80 $\pm$ 10 K., a very low
temperature to be associated with a H II region. This temperature is
more likely associated with thermal dust emission.

Possibly, the high mass accretion rates of the
infalling material associated with W51 North ($\sim$ 10$^{-3}$
$M_\odot$ yr$^{-1}$) could be likely quenching or trapping the
development of an ultra-compact HII region, so that the free-free
emission from the ionized material is undetectable at centimeter
wavelengths, see for a reference of these phenomena:
\citet{oso1999,ke2003}.  Another possibly is we have a quite young
massive protostar that has not developed an HII region yet.

\begin{figure}[!ht]
\begin{center} 
\includegraphics[scale=0.56, angle=0]{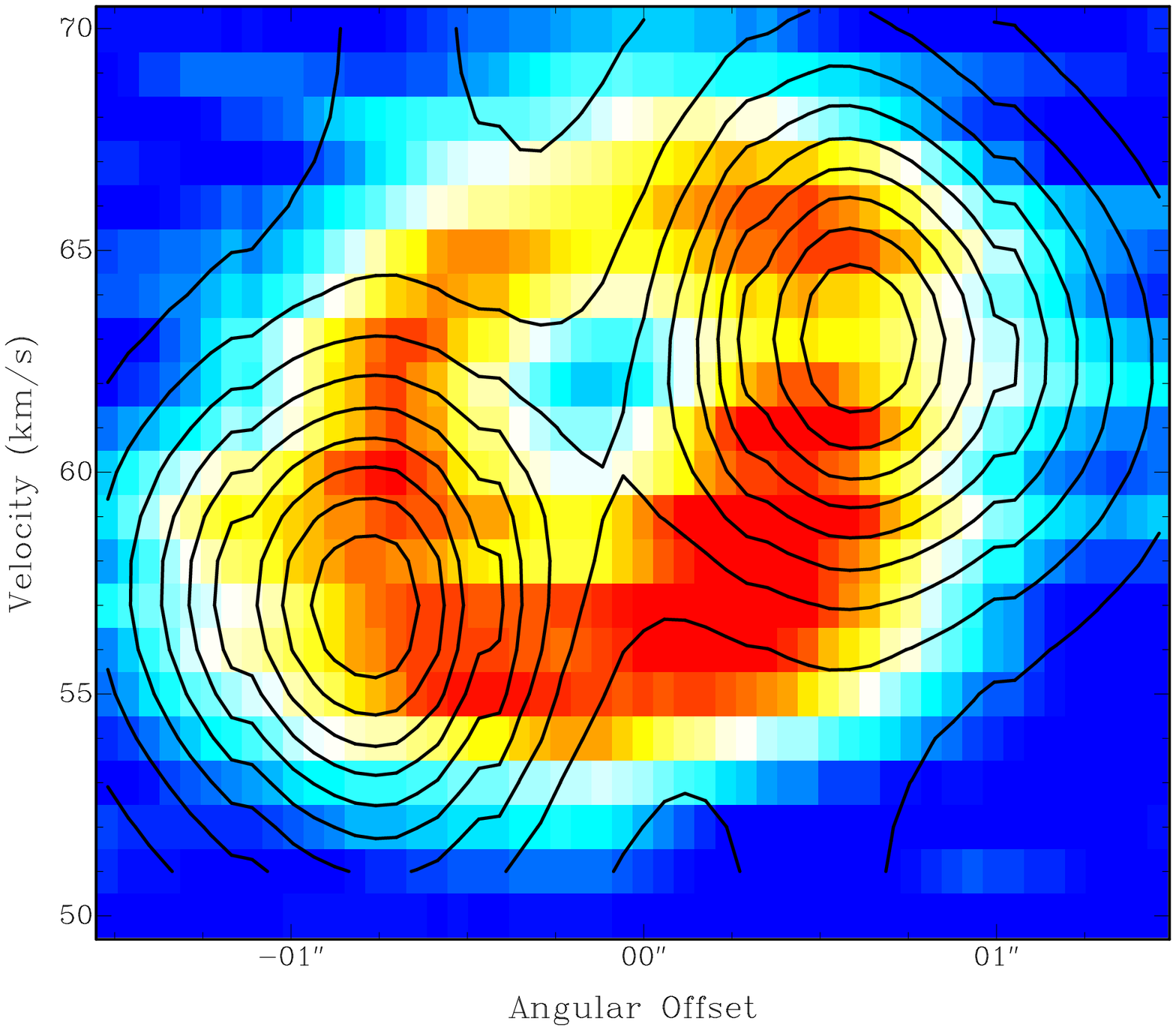}
\caption{\scriptsize 
Position-velocity diagram of the molecular ring computed at P.A. = 22$^\circ$ (colors)
overlaid with the position-velocity diagram from our model (contours).  
The contours are from 10\%
to 90\% with steps of 10\% of the peak of the line emission of our LTE model. 
The units of the vertical axis are in arcseconds.
The systemic LSR radial velocity of the ambient molecular
cloud is about 60.0 km s$^{-1}$.  The synthesized beam of the line
image is 0.58$''$ $\times$ 0.43$''$ with a P.A. of 54.8$^\circ$. 
The spectral resolution was smoothed to 1 km s$^{-1}$. }
\end{center}
\end{figure}

In Figure 1, we also show the morphology of the continuum source at
1.3 mm. At both wavelengths, the deconvolved sizes are similar,
with a size on the order 3500 $\times$ 1500 AU, see Table 1.
Certainly, a very large disk compared with those observed in low-mass stars ($\sim$ 100 AU), 
but with similar sizes to those observed
in early B-type protostars \citep[1000-2000 AU, e.g.][]{rica1999,sche2001,schr2006,pa2005,rod2007}.

 Finally, assuming that the emission is optically thin, a dust
 temperature value of 200 K \citep{zha1998}, a grain emissivity
 spectral index $\beta$=1 (see Figure 2), a value of the dust
 absorption coefficient that goes as $[\frac{\kappa_{\rm \nu}}{cm^2
 g^{-1}}]=0.1 [\frac{\nu}{1000 GHz}]^\beta$ and a distance to W51
 North of 6 kpc, we estimated a gas mass of the dusty disk of 40
 M$_\odot$. This value is consistent with the mass (100 M$_\odot$)
 obtained for the large dusty structures found in W51 North by
 \citet{zha1998, zap2008} .

\subsection{A collimated molecular outflow}

The SiO[5$\rightarrow$4] zero moment map shows the presence of a
compact and collimated ($\sim$ 14$^\circ$) north-south bipolar
molecular outflow. The redshifted radial velocities are from +60 to
+95 km s$^{-1}$ and the blueshifted ones are from +20 to +58 km
s$^{-1}$; the systemic LSR radial velocity of the ambient molecular
cloud is about at 59 km s$^{-1}$.  The receding radial velocities
(redshifted) of the outflow are located toward the north, while the
approaching radial velocities (blueshifted) are toward the
southwest. The outflow emanates from the mm source imaged at 1.3 and 7
mm wavelengths (see Figure 1).

In Figure 1, we have also overlaid the SiO[5$\rightarrow$4] molecular
emission from the bipolar outflow with the positions of the blue- and
red-shifted water maser spots reported by \citet{ein2002}.  It is
evident how the water maser spots trace the innermost parts of the
SiO[5$\rightarrow$4] molecular bipolar outflow as observed on many
other molecular outflows and proposed for this case by
\citet{zap2008}.

\begin{figure}[!ht]
\begin{center} 
\includegraphics[scale=0.21, angle=0]{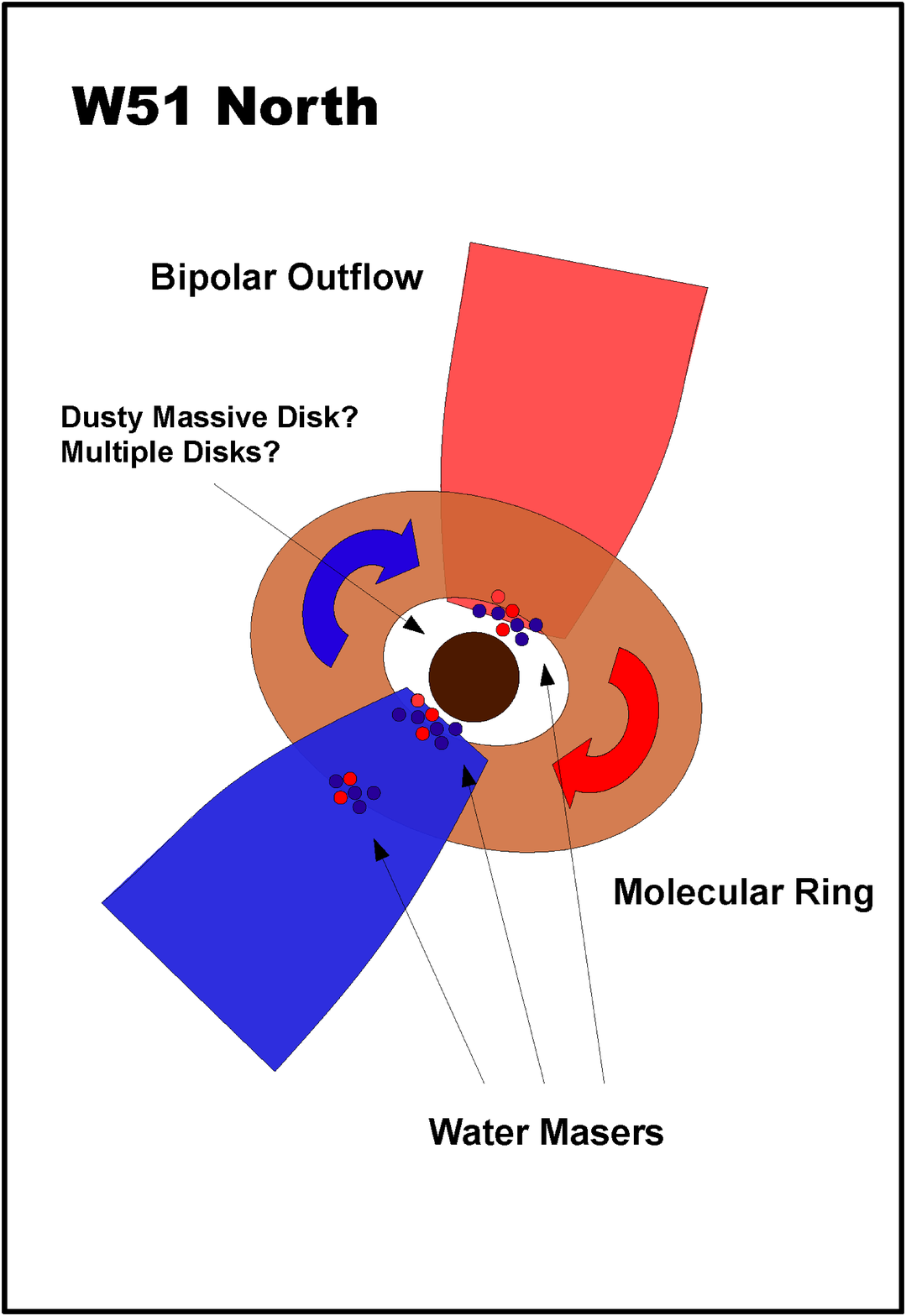}
\caption{\scriptsize Artist's  conception of the molecular ring, 
the bipolar outflow and
the central massive disk found in the high-mass protostar W51 North.}
\end{center}
\vspace{0.5cm}
\end{figure}

In the other hand, the SiO masers reported also in
\citet{ein2002} and associated with very luminous
high-mass star forming regions \citep{zap2008b}, might be tracing even
more dense parts of the outflow at very small scales, as has been
observed for the massive protostar {\it Source I} located in the
closest massive star forming region, Orion \citep{me1995,re2007}. We
have marked the position of the SiO maser source in our Figure 1. The
P.A. of the SiO[5$\rightarrow$4] molecular outflow is 150$^\circ$
$\pm$ 20$^\circ$, almost perpendicular to the orientation of the dusty
source, as mentioned before. The large radial velocities of the
outflow and the morphology of the SiO[5$\rightarrow$4] molecular
emission suggest that it is nearly perpendicular to the plane of the
sky.  This is also suggested by the location of water maser radial
redshifted and blueshifted velocities, see
\citet{imaietal2002,ein2002}.

Assuming that we are in LTE and the molecular emission is optically
thin, we can estimate the mass of the outflow for the SiO molecule in
the transition $\Delta$J= $5 \rightarrow 4$ from the equation:

\begin{equation}
M_{H_2}=1.75 \times 10^{31} \frac{X_{H_2}}{X_{SiO}} \frac{D^2}{\nu^2}
\frac{ \exp \left ( \frac{20.1}{T_{ex}} \right ) \int S_{\nu} dV
\Delta \Omega}{ \left (1 - \exp \left[\frac{-10.5}{T_{ex}} \right ]
\right ) }
\end{equation} 

\noindent
where $ \frac{X_{H_2}}{X_{SiO}}$, is the fractional abundance between
the silicon monoxide and the molecular hydrogen, that for this case we
took a value of 10$^{7}$, D is distance to W51 North, $ \Delta \Omega$
is solid angle, $\nu$ the rest frequency of the line, S$_{\nu}$ the
flux density and T$_{ex}$ the excitation temperature equal 50 K. We
estimated a mass for the outflow of 200 M$_\odot$.  It is important to
mention that the fractional abundance between the silicon monoxide and
the molecular hydrogen varies in every star forming regions (SFRs) and
thus the mass of the outflow can be over/underestimated. The value for
the fractional abundance assumed here seems to be consistent for a few
SFRs, see \citet{zi1987,mi1992,zha1995}.

For a mechanical force of F$_M$=200 M$_\odot$ 30 km s$^{-1}$/ 1800
yr$^{-1}$=3.3 M$_\odot$ km s$^{-1}$ yr$^{-1}$ and from the correlation
presented in \citet{wu2004} for the outflow mechanical force versus
the bolometric luminosity of the exciting source, we very roughly
estimated a luminosity for the central powering source of more 10$^5$
L$_\odot$ which corresponds to a massive early O-type protostar. This
spectral type for the central star is in good agreement with that
obtained from the dynamical considerations.

\subsection{A hot molecular gas ring}

A large and flattened molecular ring (with an inner cavity of about
3000 AU) surrounding the dusty disk was traced by the emission of the
molecule SO$_2$[22$_{2,20}$ $\rightarrow$ 22$_{1,21}$] (see Figure
1). 

The ring has a deconvolved radius of order 6000 AU with an orientation
(P.A.= 22$^\circ$) almost perpendicular to that of the molecular
outflow and similar to the presumed dusty disk. In Figure 3, we show
the position-velocity diagram of the
molecular SO$_2$[22$_{2,20}$ $\rightarrow$ 22$_{1,21}$] emission from the
ring structure.
From this figure, one can see a ring structure with a modest
velocity gradient across the ring of $\sim$ 5 km s$^{-1}$ over
2.0\arcsec, produced by rotation and infall. The redshifted radial
velocities located to the east, while the blueshifted ones to the
west.  The small velocity gradient is likely due to the small
inclination angle of ring with respect to plane of sky. Similar rings
velocity structures were also found traced by the emission of other
molecules species, {\it e.g} HC$_3$N, H$_2$CO, NH$_2$CHO and CH$_3$OH.
 
Very recently, however, \citet{zap2008} reported a large structure or
envelope with a size of 4 $\times$ 10$^4$ AU and with a P.A. =
90$^\circ$ traced by the C$_2$H$_3$CN molecule and centered in W51
North.  This molecular structure seems rotate in an opposite direction
(with the blueshifted velocities to the east while the redshifted ones
to the east) to the molecular compact ring traced by the
SO$_2$[22$_{2,20}$ $\rightarrow$ 22$_{1,21}$] . As discussed in
\citet{zap2008}, other molecules such as HCOOCH$_3$ and CH$_3$OH at
same scales associated with W51 North were very much contaminated by
the emission from the hot molecular core associated with W51d2 do not
allowing to confirm the velocity shift.  We thus think that the
C$_2$H$_3$CN molecular core in W51 North was also contaminated with
emission coming from W51d2 given therefore the impression of rotation.

We are more confident of the velocity gradient across the molecular
ring are correct because we have resolved the molecular source and
other molecules show similar velocity structures.

In Figure 4 and 5, we additionally present the velocity channel maps and the
position-velocity diagram of the
SO$_2$[22$_{2,20}$ $\rightarrow$ 22$_{1,21}$] emission from the ring,
overlaid with a simple LTE model of a large infalling Keplerian ring
with an inner cavity with sizes (of the ring and the inner cavity)
similar to those observed.

We assumed in this model that the thermal molecular line emission of
the ring is in local thermodynamic equilibrium, and took from the
literature some know values as the excitation temperature, the density
and the infall velocity \citep[see][]{zha1998,zap2008}.  The best fit
was found recurrently until we obtained similar structures
in our model to those imaged (see, Table 1 and Figures 3 and 4).  
The model fits the observations well.  

This model, in addition, restricted the position
of the ring on the sky. It is almost in face-on, with an inclination
angle with respect to plane of the sky less than 30$^\circ$ and a
P.A. equal to 22$^\circ$. It is important to mention that if 
the inclination angle is smaller, the dynamical mass will increase by much more, 
with a reason inversely proportional to the sin$^2$({\it i}), where {\it i} is the
inclination angle. Thus, the central object could have a stellar mass of 
more than 60 M$_\odot$.

It is interesting to note that the mass of the disk (40 M$_\odot$)
might be on the order of the stellar mass 
($\geq$ 60 M$_\odot$). This has seen before for the massive young disks associated
with the high-mass protostars located in the star forming region:
NGC6334N(I) \citep{rod2007}.

The nature of large inner cavity in the molecular ring may be due 
to opacity effects, photodissociation of the molecules, or 
clearing of the inner disk either from a multiple 
system of compact circumstellar disks already formed in the center 
of the ring or by the outflow itself. If the nature of the ring
is because of  opacity effects or photodissociation of the molecules 
very close to the central object suggesting that both the molecular 
ring and circumstellar disk could be part of a same extended and flattened 
structure. Observations with a much better angular resolution of high-density tracers
and the continuum at millimeter wavelengths may determinate the nature 
of the inner cavity in the ring.

The strong feature observed toward the east of the ring might be
originated from the interaction of the outflow with the ring itself.
However, more observations at high angular resolution with different
molecular probes sensitive to different cloud properties will be quite
important for studying the higher excitation material within the ring.

In Figure 5 we show artist's  conception of the molecular ring, 
the bipolar outflow and the central massive disk found 
in the high-mass protostar W51 North.

\section{Summary}

W51 North presents a promising laboratory for future
studies on formation of very high-mass stars. We have
observed this young massive protostar at spatial scales of 0.4$''$ 
using the SMA and VLA and found the following:

\begin{itemize}

\item A massive (40 M$_\odot$) and large ($\sim$ 3000 AU) dusty circumstellar
disk around a single high-mass protostar or a compact small stellar system.

\item A hot and rotating molecular ring with a inner cavity of about 3000 AU
and a size of about 9,000 AU.

\item A highly collimated and massive (200 M$_\odot$) bipolar outflow that emanates
with an orientation almost perpendicular to the circumstellar disk and the ring.  
This collimated outflow suggests the presence of a single massive protostar in
the center of the circumstellar disk.  

\item Our data reveled that H$_2$O and SiO masers associated with this highly
embedded protostar are tracing the innermost parts of the SiO[5$\rightarrow$4]
thermal bipolar outflow. 
 
\item A molecular hybrid LTE model of a
Keplerian infalling ring with an inner cavity ($\sim$ 3000 AU) and 
a stellar mass of more than 60 M$_\odot$ agree well with the
SO$_2$[22$_{2,20}$ $\rightarrow$ 22$_{1,21}$] line observations.

\end{itemize}

These results suggest that mechanisms, such as mergers 
of low- and intermediate- mass stars,
might be not necessary for forming very massive stars.

\begin{deluxetable}{ccc}
\tabletypesize{\scriptsize }
\tablecaption{Parameters for the molecular gas ring LTE model }
\tablewidth{0pt}
\centering
\tablehead{
 Name & Parameter & Value }
\startdata
Systematic Velocity & V$_{LSR} $ & 60.0 km s$^{-1}$\\ Orientation & PA
& 22$^\circ$ $\pm$ 10$^\circ$ \\ Inclination & {\it i}& $\leq$ 30$^\circ$\\ Density &
$\rho$ & 10$^8$ cm$^{-3}$\\ Inner Radius & R$_i$ & 3000 AU $\pm$ 200 AU \\ Outer
Radius & R$_o$ & 6000 AU $\pm$ 200 AU \\ Temperature of the central object & T & 500 K\\
Dust Exponent & $\beta$ & 1.0\\ Distance & D & 6000 pc\\ Reference
radius & r & 6000 AU \\ Power law index density & $\alpha$ & 2.75 \\
Kinetic temperature & & \\ at referenced radius & T$_{kin}$ & 200 K \\
Excitation temperature & & \\ at referenced radius & T$_{exc}$ & 200 K
\\ Power law index T$_{kin}$ & $\gamma$ & 0.6 \\ Power law index
T$_{exc}$ & $\delta$ & 0.6 \\ Scale Height of disk & H & 60 AU $\pm$ 20 AU \\
Dynamical mass & M & $\geq$ 100 M$_\odot$\\ infall velocity & V$_{inf}$
& 4 km s$^{-1}$
\enddata
\end{deluxetable}



\acknowledgments
We thank all the SMA and NRAO staff members for making these
observations possible.  The Submillimeter Array (SMA) is a joint
project between the Smithsonian Astrophysical Observatory and the
Academia Sinica Institute of Astronomy and Astrophysics, and is funded
by the Smithsonian Institution and the Academia Sinica. NRAO is
operated by Associated University, Inc., under contract with the
National Science Foundation.



{\it Facilities:} \facility{ Submillimeter Array (SMA)},
\facility{Very Large Array (VLA)}.

\bibliographystyle{apj}
\bibliography{biblio}

\begin{thebibliography}{43}
\expandafter\ifx\csname natexlab\endcsname\relax\def\natexlab#1{#1}\fi

\bibitem[{{Barbosa} {et~al.}(2008){Barbosa}, {Blum}, {Conti}, {Damineli}, \&
  {Figuer{\^e}do}}]{bar2008}
{Barbosa}, C.~L., {Blum}, R.~D., {Conti}, P.~S., {Damineli}, A., \&
  {Figuer{\^e}do}, E. 2008, \apjl, 678, L55

\bibitem[{{Beltr{\'a}n} {et~al.}(2006){Beltr{\'a}n}, {Cesaroni}, {Codella},
  {Testi}, {Furuya}, \& {Olmi}}]{beltranetal2006}
{Beltr{\'a}n}, M.~T., {Cesaroni}, R., {Codella}, C., {Testi}, L., {Furuya},
  R.~S., \& {Olmi}, L. 2006, \nat, 443, 427

\bibitem[{{Beltr{\'a}n} {et~al.}(2005){Beltr{\'a}n}, {Cesaroni}, {Neri},
  {Codella}, {Furuya}, {Testi}, \& {Olmi}}]{beltranetal2005}
{Beltr{\'a}n}, M.~T., {Cesaroni}, R., {Neri}, R., {Codella}, C., {Furuya},
  R.~S., {Testi}, L., \& {Olmi}, L. 2005, \aap, 435, 901

\bibitem[{{Beuther} \& {Walsh}(2008)}]{be2008}
{Beuther}, H. \& {Walsh}, A.~J. 2008, \apjl, 673, L55

\bibitem[{{Cesaroni} {et~al.}(1999){Cesaroni}, {Felli}, {Jenness}, {Neri},
  {Olmi}, {Robberto}, {Testi}, \& {Walmsley}}]{rica1999}
{Cesaroni}, R., {Felli}, M., {Jenness}, T., {Neri}, R., {Olmi}, L., {Robberto},
  M., {Testi}, L., \& {Walmsley}, C.~M. 1999, \aap, 345, 949

\bibitem[{{Cesaroni} {et~al.}(2006){Cesaroni}, {Galli}, {Lodato}, {Walmsley},
  \& {Zhang}}]{cesaronietal2006}
{Cesaroni}, R., {Galli}, D., {Lodato}, G., {Walmsley}, M., \& {Zhang}, Q. 2006,
  \nat, 444, 703

\bibitem[{{Eisner} {et~al.}(2002){Eisner}, {Greenhill}, {Herrnstein}, {Moran},
  \& {Menten}}]{ein2002}
{Eisner}, J.~A., {Greenhill}, L.~J., {Herrnstein}, J.~R., {Moran}, J.~M., \&
  {Menten}, K.~M. 2002, \apj, 569, 334

\bibitem[{{Erickson} \& {Tokunaga}(1980)}]{Eri1980}
{Erickson}, E.~F. \& {Tokunaga}, A.~T. 1980, \apj, 238, 596

\bibitem[{{Figuer{\^e}do} {et~al.}(2008){Figuer{\^e}do}, {Blum}, {Damineli},
  {Conti}, \& {Barbosa}}]{Fig2008}
{Figuer{\^e}do}, E., {Blum}, R.~D., {Damineli}, A., {Conti}, P.~S., \&
  {Barbosa}, C.~L. 2008, \aj, 136, 221

\bibitem[{{Gaume} \& {Mutel}(1987)}]{gau1987}
{Gaume}, R.~A. \& {Mutel}, R.~L. 1987, \apjs, 65, 193

\bibitem[{{Genzel} {et~al.}(1981){Genzel}, {Downes}, {Schneps}, {Reid},
  {Moran}, {Kogan}, {Kostenko}, {Matveenko}, \& {Ronnang}}]{genzeletal1981}
{Genzel}, R., {Downes}, D., {Schneps}, M.~H., {Reid}, M.~J., {Moran}, J.~M.,
  {Kogan}, L.~R., {Kostenko}, V.~I., {Matveenko}, L.~I., \& {Ronnang}, B. 1981,
  \apj, 247, 1039

\bibitem[{{Ho} {et~al.}(1983){Ho}, {Das}, \& {Genzel}}]{ho1983}
{Ho}, P.~T.~P., {Das}, A., \& {Genzel}, R. 1983, \apj, 266, 596

\bibitem[{{Ho} {et~al.}(2004){Ho}, {Moran}, \& {Lo}}]{ho2004}
{Ho}, P.~T.~P., {Moran}, J.~M., \& {Lo}, K.~Y. 2004, \apjl, 616, L1

\bibitem[{{Imai} {et~al.}(2002){Imai}, {Watanabe}, {Omodaka}, {Nishio},
  {Kameya}, {Miyaji}, \& {Nakajima}}]{imaietal2002}
{Imai}, H., {Watanabe}, T., {Omodaka}, T., {Nishio}, M., {Kameya}, O.,
  {Miyaji}, T., \& {Nakajima}, J. 2002, \pasj, 54, 741

\bibitem[{{Jim{\'e}nez-Serra} {et~al.}(2007){Jim{\'e}nez-Serra},
  {Mart{\'{\i}}n-Pintado}, {Rodr{\'{\i}}guez-Franco}, {Chandler}, {Comito}, \&
  {Schilke}}]{jimena2007}
{Jim{\'e}nez-Serra}, I., {Mart{\'{\i}}n-Pintado}, J.,
  {Rodr{\'{\i}}guez-Franco}, A., {Chandler}, C., {Comito}, C., \& {Schilke}, P.
  2007, \apjl, 661, L187

\bibitem[{{Keto}(2003)}]{ke2003}
{Keto}, E. 2003, \apj, 599, 1196

\bibitem[{{Lacy} {et~al.}(2007){Lacy}, {Jaffe}, {Zhu}, {Richter}, {Bitner},
  {Greathouse}, {Volk}, {Geballe}, \& {Mehringer}}]{la2007}
{Lacy}, J.~H., {Jaffe}, D.~T., {Zhu}, Q., {Richter}, M.~J., {Bitner}, M.~A.,
  {Greathouse}, T.~K., {Volk}, K., {Geballe}, T.~R., \& {Mehringer}, D.~M.
  2007, \apjl, 658, L45

\bibitem[{{Leurini} {et~al.}(2007){Leurini}, {Beuther}, {Schilke}, {Wyrowski},
  {Zhang}, \& {Menten}}]{leu07}
{Leurini}, S., {Beuther}, H., {Schilke}, P., {Wyrowski}, F., {Zhang}, Q., \&
  {Menten}, K.~M. 2007, \aap, 475, 925

\bibitem[{{Mehringer}(1994)}]{Meh1994}
{Mehringer}, D.~M. 1994, \apjs, 91, 713

\bibitem[{{Menten} \& {Reid}(1995)}]{me1995}
{Menten}, K.~M. \& {Reid}, M.~J. 1995, \apjl, 445, L157

\bibitem[{{Mikami} {et~al.}(1992){Mikami}, {Umemoto}, {Yamamoto}, \&
  {Saito}}]{mi1992}
{Mikami}, H., {Umemoto}, T., {Yamamoto}, S., \& {Saito}, S. 1992, \apjl, 392,
  L87

\bibitem[{{Morita} {et~al.}(1992){Morita}, {Hasegawa}, {Ukita}, {Okumura}, \&
  {Ishiguro}}]{mor1992}
{Morita}, K.-I., {Hasegawa}, T., {Ukita}, N., {Okumura}, S.~K., \& {Ishiguro},
  M. 1992, \pasj, 44, 373

\bibitem[{{Osorio} {et~al.}(1999){Osorio}, {Lizano}, \& {D'Alessio}}]{oso1999}
{Osorio}, M., {Lizano}, S., \& {D'Alessio}, P. 1999, \apj, 525, 808

\bibitem[{{Patel} {et~al.}(2005){Patel}, {Curiel}, {Sridharan}, {Zhang},
  {Hunter}, {Ho}, {Torrelles}, {Moran}, {G{\'o}mez}, \& {Anglada}}]{pa2005}
{Patel}, N.~A., {Curiel}, S., {Sridharan}, T.~K., {Zhang}, Q., {Hunter}, T.~R.,
  {Ho}, P.~T.~P., {Torrelles}, J.~M., {Moran}, J.~M., {G{\'o}mez}, J.~F., \&
  {Anglada}, G. 2005, \nat, 437, 109

\bibitem[{{Reid} {et~al.}(2007){Reid}, {Menten}, {Greenhill}, \&
  {Chandler}}]{re2007}
{Reid}, M.~J., {Menten}, K.~M., {Greenhill}, L.~J., \& {Chandler}, C.~J. 2007,
  \apj, 664, 950

\bibitem[{{Rodr{\'{\i}}guez} {et~al.}(2007){Rodr{\'{\i}}guez}, {Zapata}, \&
  {Ho}}]{rod2007}
{Rodr{\'{\i}}guez}, L.~F., {Zapata}, L.~A., \& {Ho}, P.~T.~P. 2007, \apjl, 654,
  L143

\bibitem[{{Rudolph} {et~al.}(1990){Rudolph}, {Welch}, {Palmer}, \&
  {Dubrulle}}]{rud1990}
{Rudolph}, A., {Welch}, W.~J., {Palmer}, P., \& {Dubrulle}, B. 1990, \apj, 363,
  528

\bibitem[{{Sandell} {et~al.}(2003){Sandell}, {Wright}, \&
  {Forster}}]{sandelletal2003}
{Sandell}, G., {Wright}, M., \& {Forster}, J.~R. 2003, \apjl, 590, L45

\bibitem[{{Schilke} {et~al.}(1997){Schilke}, {Walmsley}, {Pineau des Forets},
  \& {Flower}}]{sch97}
{Schilke}, P., {Walmsley}, C.~M., {Pineau des Forets}, G., \& {Flower}, D.~R.
  1997, \aap, 321, 293

\bibitem[{{Schneps} {et~al.}(1981){Schneps}, {Lane}, {Downes}, {Moran},
  {Genzel}, \& {Reid}}]{Schnepsetal1981}
{Schneps}, M.~H., {Lane}, A.~P., {Downes}, D., {Moran}, J.~M., {Genzel}, R., \&
  {Reid}, M.~J. 1981, \apj, 249, 124

\bibitem[{{Schreyer} {et~al.}(2006){Schreyer}, {Semenov}, {Henning}, \&
  {Forbrich}}]{schr2006}
{Schreyer}, K., {Semenov}, D., {Henning}, T., \& {Forbrich}, J. 2006, \apjl,
  637, L129

\bibitem[{{Shepherd} {et~al.}(2001){Shepherd}, {Claussen}, \&
  {Kurtz}}]{sche2001}
{Shepherd}, D.~S., {Claussen}, M.~J., \& {Kurtz}, S.~E. 2001, Science, 292,
  1513

\bibitem[{{Sollins} \& {Ho}(2005)}]{sollinsetal2005}
{Sollins}, P.~K. \& {Ho}, P.~T.~P. 2005, \apj, 630, 987

\bibitem[{{Sollins} {et~al.}(2004){Sollins}, {Zhang}, \& {Ho}}]{soll2004}
{Sollins}, P.~K., {Zhang}, Q., \& {Ho}, P.~T.~P. 2004, \apj, 606, 943

\bibitem[{{van Dishoeck} \& {Blake}(1998)}]{van98}
{van Dishoeck}, E.~F. \& {Blake}, G.~A. 1998, \araa, 36, 317

\bibitem[{{Wu} {et~al.}(2004){Wu}, {Wei}, {Zhao}, {Shi}, {Yu}, {Qin}, \&
  {Huang}}]{wu2004}
{Wu}, Y., {Wei}, Y., {Zhao}, M., {Shi}, Y., {Yu}, W., {Qin}, S., \& {Huang}, M.
  2004, \aap, 426, 503

\bibitem[{{Xu} {et~al.}(2008){Xu}, {Reid}, {Menten}, {Brunthaler}, {Zheng}, \&
  {Moscadelli}}]{xu2008}
{Xu}, Y., {Reid}, M.~J., {Menten}, K.~M., {Brunthaler}, A., {Zheng}, X.~W., \&
  {Moscadelli}, L. 2008, ArXiv e-prints

\bibitem[{{Zapata} {et~al.}(2008{\natexlab{a}}){Zapata}, {Menten}, {Reid}, \&
  {Beuther}}]{zap2008b}
{Zapata}, L.~A., {Menten}, K., {Reid}, M., \& {Beuther}, H. 2008{\natexlab{a}},
  ArXiv e-prints

\bibitem[{{Zapata} {et~al.}(2008{\natexlab{b}}){Zapata}, {Palau}, {Ho},
  {Schilke}, {Garrod}, {Rodr{\'{\i}}guez}, \& {Menten}}]{zap2008}
{Zapata}, L.~A., {Palau}, A., {Ho}, P.~T.~P., {Schilke}, P., {Garrod}, R.~T.,
  {Rodr{\'{\i}}guez}, L.~F., \& {Menten}, K. 2008{\natexlab{b}}, \aap, 479, L25

\bibitem[{{Zhang} \& {Ho}(1997)}]{zha1997}
{Zhang}, Q. \& {Ho}, P.~T.~P. 1997, \apj, 488, 241

\bibitem[{{Zhang} {et~al.}(1998){Zhang}, {Ho}, \& {Ohashi}}]{zha1998}
{Zhang}, Q., {Ho}, P.~T.~P., \& {Ohashi}, N. 1998, \apj, 494, 636

\bibitem[{{Zhang} {et~al.}(1995){Zhang}, {Ho}, {Wright}, \& {Wilner}}]{zha1995}
{Zhang}, Q., {Ho}, P.~T.~P., {Wright}, M.~C.~H., \& {Wilner}, D.~J. 1995,
  \apjl, 451, L71+

\bibitem[{{Ziurys} \& {Friberg}(1987)}]{zi1987}
{Ziurys}, L.~M. \& {Friberg}, P. 1987, \apjl, 314, L49

\end{thebibliography}

\end{document}